\documentclass[conference]{IEEEtran}
\IEEEoverridecommandlockouts
\usepackage{cite}
\usepackage{amsmath,amssymb,amsfonts}
\usepackage{algorithmic}
\usepackage{graphicx}
\usepackage{textcomp}
\usepackage{xcolor}
\usepackage{subcaption}
\usepackage{hyperref}
\makeatletter
\def\UrlAlphabet{%
      \do\a\do\b\do\c\do\d\do\e\do\f\do\g\do\h\do\i\do\j%
      \do\k\do\l\do\m\do\n\do\o\do\p\do\q\do\r\do\s\do\t%
      \do\u\do\v\do\w\do\x\do\y\do\z\do\A\do\B\do\C\do\D%
      \do\E\do\F\do\G\do\H\do\I\do\J\do\K\do\L\do\M\do\N%
      \do\O\do\P\do\Q\do\R\do\S\do\T\do\U\do\V\do\W\do\X%
      \do\Y\do\Z}
\def\UrlDigits{\do\1\do\2\do\3\do\4\do\5\do\6\do\7\do\8\do\9\do\0}
\g@addto@macro{\UrlBreaks}{\UrlOrds}
\g@addto@macro{\UrlBreaks}{\UrlAlphabet}
\g@addto@macro{\UrlBreaks}{\UrlDigits}
\makeatother

\def\BibTeX{{\rm B\kern-.05em{\sc i\kern-.025em b}\kern-.08em
    T\kern-.1667em\lower.7ex\hbox{E}\kern-.125emX}}
    
\begin{document}

\title{Distributed PMCW Radar Network in Presence of Phase Noise\\
{\footnotesize }
\thanks{}
}
\author{\IEEEauthorblockN{Jialun Kou\textsuperscript{*}, Marc Bauduin\textsuperscript{†}, André Bourdoux\textsuperscript{†}, Sofie Pollin\textsuperscript{*†}}
\IEEEauthorblockA{\textit{\textsuperscript{*}Department of Electrical Engineering (ESAT), KU Leuven, Belgium } \\
\textit{\textsuperscript{†}Interuniversity Microelectronics Centre (IMEC), Leuven, Belgium}\\
\{jialun.kou, sofie.pollin\}@kuleuven.be, \{marc.bauduin, andre.bourdoux\}@imec.be\\
}
}

\maketitle

\begin{abstract}
In Frequency Modulated Continuous Waveform (FMCW) radar systems, the phase noise from the Phase-Locked Loop (PLL) can increase the noise floor in the Range-Doppler map. The adverse effects of phase noise on close targets can be mitigated if the transmitter (Tx) and receiver (Rx) employ the same chirp, a phenomenon known as the range correlation effect.

In the context of a multi-static radar network, sharing the chirp between distant radars becomes challenging. Each radar generates its own chirp, leading to uncorrelated phase noise. Consequently, the system performance cannot benefit from the range correlation effect.

Previous studies show that selecting a suitable code sequence for a Phase Modulated Continuous Waveform (PMCW) radar can reduce the impact of uncorrelated phase noise in the range dimension. In this paper, we demonstrate how to leverage this property to exploit both the mono- and multi-static signals of each radar in the network without having to share any signal at the carrier frequency. The paper introduces a detailed signal model for PMCW radar networks, analyzing both correlated and uncorrelated phase noise effects in the Doppler dimension. Additionally, a solution for compensating uncorrelated phase noise in Doppler is presented and supported by numerical results.
\end{abstract}
\begin{IEEEkeywords}
PMCW radar, Phase noise, Radar network, Multi-static radar 
\end{IEEEkeywords}

\section{Introduction}
Radar networks offer many advantages in complex scenarios, providing enhanced coverage, improved accuracy, and flexibility. In a Frequency Modulated Continuous Waveform (FMCW) radar network, the chirp is susceptible to non-idealities such as phase noise (PN), leading degradations in the Range-Doppler map. In the mono-static scenario, where received and transmitted signals come from the same radar, the impact of PN is attenuated for close targets. This is called the range correlation effect \cite{b8}. However, in the multi-static scenario, where signals come from different radars, each equipped with a distinct Phase-Locked Loop (PLL), the PN introduced by each PLL is uncorrelated. Consequently, this uncorrelated PN cannot be attenuated by range correlation, resulting in an increased noise floor. Achieving effective range correlation in multi-static setups requires a fully coherent radar network, often necessitating complex hardware designs \cite{b9}.

A possible solution to bypass the need for complicated hardware synchronization is to implement a radar-repeater network \cite{b10}. In such a network, there is only one transceiver, and repeaters do not generate new waveforms but instead repeat the received signal back to free space, producing a bi-static target response. As all signals originate from the same waveform generator, the entire network is fully coherent. However, in the presence of multiple targets, the repeater may produce ghost bi-static target responses \cite{b11}.

In this study, we leverage waveform design to provide robustness to PN. Study \cite{b3} demonstrates that Phase Modulated Continuous Waveform (PMCW) radar can effectively reduce the impact of PN in the range profile by selecting a suitable code sequence. In addition, \cite{b12} analyzes the effect of uncorrelated PN on the PMCW radar Range-Doppler map in the context of mutual interferences scenario. It shows that uncorrelated PN can produce a ridge in the Doppler dimension but does not offer a solution for compensating uncorrelated PN in slow time. Another paper, \cite{b13}, presents a solution to compensate for the impact of PN along slow time, improving Doppler profiles. However, it relies on a calibration target. 

In this paper, we exploit the property of PMCW radar, where PN produces low-range sidelobes, to design a multi-static radar network. We present a comprehensive signal model for PMCW radar networks, analyzing the impact of both correlated and uncorrelated PN. Additionally, we use Line-of-Sight (LOS) propagation between the two radars to effectively compensate for uncorrelated PN along the slow time. As the proposed solution can reduce the impact of uncorrelated PN, the synchronization requirements between the different radars in the network can be relaxed. In this paper, we assume that the two radars are time-synchronized (same start-of-frame) and share a low frequency Local Oscillator (LO). However, they do not share the same PLL, resulting in uncorrelated PN between the two radars.

The outline of this paper is as follows: In Section II, we introduce the ideal system model and PN model. Section III presents the mono-static and bi-static signal models with PN. Section IV details the solution to compensate for uncorrelated PN, and Section V provides numerical results.

\section{System model}
\subsection{PMCW signal model} \label{section A}
The PMCW waveform can be characterized as:
\begin{align}
s(t) &= \sum_{l=0}^{L_c-1} b(l) \text{rect}\left(\frac{t-lT_c}{T_c}\right), \quad 0 \leq t \leq T
\label{eq:single_PMCW_Waveform}
\end{align}
where $b(l)$ is the code sequence element, $L_c$ is the length of the code sequence, $T_c$ is the chip duration. $T = L_cT_c$ is the duration of the PMCW waveform, and the function $\text{rect}(x)$ is defined as:
\begin{equation}
\text{rect}(x) =
\begin{cases}
1, & \text{if } 0 \leq x \leq 1 \\
0, & \text{otherwise}
\end{cases}
\end{equation}

In this paper, the Almost Perfect Autocorrelation Sequence (APAS) is considered \cite{b6}. This is a binary code sequence with a periodic autocorrelation function that exhibits a positive peak for lags equal to 0 and a negative peak for lags at $L_c/2$. For all other lags, the autocorrelation value is exactly 0. This implies that the maximum non-ambiguous range profile is limited to $L_c/2$ instead of $L_c$ \cite{b6}.

We assume the transmission of $N$ bursts of the PMCW waveform. The signal \( S_{PM}(t) \) is defined as the periodic repetition of \( s(t) \) from \eqref{eq:single_PMCW_Waveform}, repeated $N$ times.
\begin{align} \label{eq:PMCW_Waveform}
     S_{PM}(t) = \sum_{n=0}^{N-1} s(t - nT)  \quad 0 \leq t \leq NT
\end{align}
where $n$ is the slow time index. The signal is then up-converted by the reference signal with a center frequency of $f_c$ and transmitted through the Tx antenna. In this subsection, we consider an ideal case without PN. The transmitted signal is represented as:
\begin{align} \label{eq:PMCW_TX}
S_T(t) & = S_{PM}(t) S_{ref}(t) =S_{PM}(t) e^{j2\pi f_c t} 
\end{align}
We consider a single target positioned at a distance $R$ from the radar and moving with a velocity $v$. The round-trip propagation delay is $\tilde \tau_{M}(t) = \frac{2(R+vt)}{c}= \tau_{M} + \frac{2vt}{c}$, where $c$ is the speed of light. The reflected signal is received by the Rx antenna, and the received signal is then down-converted by multiplying it with the coherent reference signal:
\begin{align}\label{eq:PMCW_RX}
S_R(t) & = S_T(t-\tilde \tau_{M}(t))S^{*}_{ref}(t) \approx S_{PM}(t-\tau_{M})e^{-j2\pi f_D t} 
\end{align}
where \(f_D = \frac{2v}{\lambda}\), and \(\ast\) represents the complex conjugate of the signal. In \eqref{eq:PMCW_RX}, we neglect the range-migration effect, the constant phase term, the thermal noise term, as well as the attenuation factor due to propagation. The received signal is subsequently sampled at intervals of \(T_c\) by the ADC.

The Range-slow-time matrix is then obtained through the periodic correlation process between \eqref{eq:PMCW_RX} and \eqref{eq:PMCW_Waveform}.

\begin{align}\label{eq:PMCW_Range_profile}
     R(p,n) & = \sum_{l=0}^{L_c-1} (b(l)b(l-\tau_d+ p))e^{-j2\pi f_D((l + p)T_c+nT)}
\end{align}
where $p$ is the index of the range bin, and $\tau_d = \lfloor \frac{\tau_M}{T_c} \rfloor$ is the propagation delay shift on the sequence. Notably, when a target is in motion, the phases of the chips in a sequence will change due to the Doppler shift. This degrades the code sequence properties, resulting in range sidelobes. These sidelobes can be compensated for by the solution proposed in \cite{b1}.

\subsection{PN model}\label{section:PN}
In real life, the LO and PLL used to up-convert the transmitted signal to the carrier frequency introduce non-idealities to the reference signal, which is known as PN, denoted as \(\phi(t)\). PN affects the ideal reference signal by introducing a phase modulation term represented as \(e^{j\phi(t)}\). The study in \cite{phase_noise} explains that, when a low-quality LO is used, the close-in PN is determined by the LO. However, when a high-quality LO is employed, the PN is only determined by the PLL. In this paper, we assume a high-quality LO is used.

The reference signal that is affected by PN can be represented as:
\begin{equation}
    \tilde S_{ref}(t) = S_{ref}(t)e^{j\phi(t)}
\end{equation} \label{eq:signal with PN}
The PN $\phi(t)$ can be represented using a piece-wise linear model, which can be generated in the frequency domain, \cite{b2}.

\begin{align}\label{eq:PN_def}
\phi(t) & = \sum_k \alpha_k \cos(2\pi f_k t + \theta_k) 
\end{align}
where \(k\) represents the frequency bin index, \(f_k = k\Delta f\) is the frequency offset, and \(\Delta f\) is determined by the inverse of the total signal duration that is simulated. \(\alpha_k\) is determined by the spectrum mask shown in Fig.~\ref{fig:psd_PN}, and \(\theta(k)\) is a random phase component with a uniform distribution between 0 and \(2\pi\). With the PN defined as \eqref{eq:PN_def}, phase modulation $e^{j\phi(t)}$ can be expressed as:
\begin{equation} 
    e^{j\phi(t)} =  \prod_{k} e^{j \alpha_k \cos(2\pi f_k t + \theta_k)}
\end{equation}
when the $\vert \alpha_k \vert $ is sufficiently small.  A linear approximation can be applied, as discussed in \cite{range_corr}.
\begin{align}
    e^{j\phi(t)} \approx 1 + j\phi(t) \label{eq:simplified_PN}
\end{align}

\section{Impact of phase noise on PMCW radar}
\subsection{Mono-static PMCW signal with correlated PN}
If we replace $S_{ref}(t) $ with $\tilde S_{ref}(t)$ in \eqref{eq:PMCW_RX}, we obtain:
\begin{align}\label{eq:PN_PMCW_RX}
  \tilde S_R(t) &= S_R(t) e^{j\Delta\Phi(t)}  
\end{align} 
Since $\phi(t-\tau_M)$ and $\phi(t)$ are coherent, range correlation will occur as $\Delta\Phi(t) = e^{j(\phi(t-\tau_M)-\phi(t))}$ \cite{range_corr}. It is important to note that, due to the high ADC sampling rate of a PMCW radar, the fast-time frequency step is quite large. Consequently, the PN attenuation due to range correlation is negligible in the range profile \cite{b3}. 

After the periodic correlation process between \eqref{eq:PN_PMCW_RX} and \eqref{eq:PMCW_Waveform}, we get the Range-slow-time matrix:
\begin{align}\label{eq:PN_PMCW_Range_profile}
    & \tilde R(p,n)   = \sum_{l=0}^{L_c-1} (b(l)b(l-\tau_d+ p))\\
    \nonumber & e^{-j2\pi f_D((l + p)T_c+nT) }e^{j\Delta\Phi((l + p)T_c+nT) }
\end{align}
Since the PN samples are not correlated with the transmitted code sequence, the sidelobe resulting from the correlated PN in the range profile is mainly attenuated through periodic correlation processing. As a result, a white Gaussian-like noise floor is formed in the range profile\cite{b3}.

We saw in \cite{b3}\cite{b12} that the range sidelobes are very low. But \cite{b12} also showed that a ridge appears in the Doppler profile. Therefore, it is required to investigate what is happening in the target range bin index along slow-time. The slow-time profile in the mono-static target range bin $\tau_d$ can be represented as:
\begin{align}\label{eq:PN_PMCW_doppler}
    & \tilde R(\tau_d,n)  \approx  \sum_{l=0}^{L_c-1} e^{-j2\pi f_D((l + \tau_d)T_c+nT) } \\
    \nonumber & ( 1+j\Delta\Phi((l + \tau_d)T_c+nT))  
\end{align}
To mathematically describe the impact of correlated PN along Doppler, we examine the term $j\Delta\Phi((l + \tau_d)T_c+nT)$.
\begin{align}
    & j\Delta\Phi((l + \tau_d)T_c+nT)  \\
    \nonumber & = j \left( \phi((l+\tau_d)T_c+nT-\tau_{M})-\phi((l+\tau_d)T_c+nT) \right) 
\end{align}    
Since $T_c$ is very short, the change in PN within interval $T_c$ is negligible, so the approximation $\phi(\tau_d T_c) = \phi(\tau_M)$ is made.
\begin{align}
    & j\Delta\Phi((l + \tau_d)T_c+nT)  \\ 
    \nonumber & \approx j \left( \phi(lT_c+nT)-\phi(lT_c+nT+\tau_M) \right) \\
   \nonumber & = \sum_{k=0}^{N-1} j\alpha_k (\cos(a-b) - \cos(a))\\
    \nonumber & = \sum_{k=0}^{N-1} j\alpha_k 2\sin(\frac{b}{2})\sin(a-\frac{b}{2})\\
    \nonumber & =\sum_{k=0}^{N-1} \alpha_k \sin(\frac{b}{2}) (e^{j(a-\frac{b}{2})}-e^{-j(a-\frac{b}{2})})
\end{align}
where $a = 2\pi f_k (lT_c+nT+\tau_M) + \theta_k$, $b = 2\pi f_k \tau_M$, $f_{k}=k \Delta f$, and the Doppler frequency resolution is $ \Delta f = \frac{1}{L_cT_cN}$, Finally, \eqref{eq:PN_PMCW_doppler} can be represented as:
\begin{align} \label{eq:range_corr}
    & \tilde R(\tau_d,n)  \approx \sum_{l=0}^{L_c-1} e^{-j2\pi f_D(lT_c+nT + \tau_{M}) } \\
    \nonumber & \biggl( 1 +
    \sum_{k=0}^{N-1} \alpha_k \sin(\pi f_{k} \tau_{M}) \left( e^{j(2\pi f_{k}(lT_c+nT + \frac{\tau_{M} }{2}) + \theta_k) } \right. \\
    \nonumber & \left. - e^{-j(2\pi f_{k}(lT_c+nT + \frac{\tau_{M} }{2})+ \theta_k)}\right)\biggl)
\end{align}

In the equation above, we observe that the slow-time profile in the mono-static signal range bin comprises both an ideal Doppler frequency shift signal and an additional PN component. The magnitude of the PN, denoted by $\alpha_k$, along the Doppler frequency is attenuated by a factor $\sin(\pi f_{k} \tau_{M})$. However, this attenuation is effective only when the condition $f_{k} \tau_{M} \leq \frac{1}{6}$ is satisfied \cite{b8}. Therefore, as the mono-static propagation delay increases, the effectiveness of PN attenuation along the Doppler dimension decreases.

\subsection{Bi-static and LOS PMCW signal with uncorrelated PN}\label{section:uncorrelated pn}
Bi-static signal propagation between the two PMCW radars involves transmitting signals from one radar system and receiving them with the other. In this section, we consider two PMCW radars configured with the same system parameters. Both radars share the low-frequency LO clock signal to mitigate carrier frequency offset, and their start-of-frame signals are synchronized, enabling simultaneous transmission of PMCW waveforms into free space.

Since the two radar systems operate with different PLLs, the PN is determined solely by the PLL, as explained in Section \ref{section:PN}, the PN introduced by each radar system is fully uncorrelated. The reference signals are then given by:
\begin{align}
    S_{\text{ref},I1}(t) & = e^{j2 \pi f_c t}e^{j\phi_{I1}(t)}\\
    S_{\text{ref},I2}(t) & = e^{j2 \pi f_c t}e^{j\phi_{I2}(t)}
\end{align}
where $\phi_{I1}(t)$ and $\phi_{I2}(t)$ represent the PN in each radar system.

The received bi-static signal can be represented as:
\begin{align} \label{eq:PN_RX_bi-static}
    & S_{R,bi}(t) = S_{PM}(t-\tilde \tau_{bi}(t))S_{ref,I1}(t-\tilde \tau_{bi}(t))S^*_{ref,I2}(t)\\
    \nonumber & \approx S_{PM}(t- \tau_{bi})e^{-j2\pi f_{D,bi} t}e^{j(\phi_{I1}(t-\tau_{bi})-\phi_{I2}(t))}    
\end{align}
where $f_{D,bi}= \frac{v_{bi}}{c} f_c$ is the Doppler frequency associated with the bi-static signal, and \( \tilde \tau_{bi}(t) = \frac{R_{bi} + v_{bi}t}{c} = \tau_{bi} + \frac{v_{bi}t}{c} \) is the bi-static propagation delay, where \( R_{\text{bi}} \) represents the bi-static range and \( v_{\text{bi}} \) is the bi-static velocity. In \eqref{eq:PN_RX_bi-static}, the same omission as in \eqref{eq:PMCW_RX} is applied.
After the periodic correlation process between \eqref{eq:PN_RX_bi-static} and \eqref{eq:PMCW_Waveform}, we get the Range-slow-time matrix: 
\begin{align} \label{eq:PN_ridge}
     & R_{bi}(p,n)  = \sum_{l=0}^{L_c-1} (b(l)b(l-\tau_{d,bi}+ p))e^{-j2\pi f_{D,bi}( (l+p)T_c + nT) }\\
    \nonumber &  e^{j(\phi_{I1}((l+p)T_c + nT -\tau_{bi})-\phi_{I2}((l+p)T_c + nT))} 
\end{align}
where $\tau_{d,bi} = \lfloor \frac{\tau_{bi}}{T_c} \rfloor$ is the bi-static propagation delay shift on the sequence.

In a scenario where two radars are mounted side by side on a single car, each radar can also receive a LOS signal from the other radar without involving any external targets.

Since the two radars do not undergo relative motion, the Doppler frequency of the LOS signal should be equal to 0. Then, the Range-slow-time matrix of the LOS signal can be represented as: 
\begin{align}\label{eq:PN_ridge_los}
& R_{los}(p,n) =  \sum_{l=0}^{L_c-1} (b(l)b(l- \tau_{d,los} + p))\\ \nonumber &
e^{j(\phi_{I1}((l+p)T_c + nT -\tau_{los})-\phi_{I2}((l+p)T_c + nT))} 
\end{align}
where $\tau_{los} = \frac{R_{los}}{c}$ is the one-way propagation delay between the two radars, and $\tau_{d,los} = \lfloor \frac{\tau_{los}}{T_c} \rfloor$ is the propagation delay shift on the sequence.
Similar to \eqref{eq:PN_PMCW_Range_profile}, the contribution of uncorrelated PN samples results in a flat noise floor in the range profile with low sidelobes.
With \eqref{eq:PN_ridge}, the slow-time profile in the bi-static target range bin $\tau_{d,bi}$ can be represented as:
\begin{align}\label{eq:PN_ridge_bi_rangebin}
      & R_{bi}(\tau_{d,bi},n)  = \sum_{l=0}^{L_c-1} e^{-j2\pi f_{D,bi}( lT_c + nT+\tau_{bi}) }\\
    \nonumber &  e^{j(\phi_{I1}(lT_c + nT)-\phi_{I2}(lT_c + nT + \tau_{bi}))}    
\end{align}
and with \eqref{eq:PN_ridge_los}, the slow-time profile in the LOS range bin $\tau_{d,los}$ can be represented as:
\begin{align}\label{eq:PN_ridge_los_rangebin}
      & R_{los}(\tau_{d,los},n)  = \sum_{l=0}^{L_c-1} e^{j(\phi_{I1}(lT_c + nT)-\phi_{I2}(lT_c + nT + \tau_{los}))}    
\end{align}
When examining the slow-time profile in the bi-static \eqref{eq:PN_ridge_bi_rangebin} and LOS \eqref{eq:PN_ridge_los_rangebin} target range bin, it becomes evident that the uncorrelated PN cannot be compensated through range correlation. Consequently, the uncorrelated PN from both the bi-static and LOS signals forms a distinctive ridge along the Doppler dimension at their respective range bins.

\section{Solution to compensate uncorrelated PN}

\subsection{Uncorrelated PN in Range profile} \label{range_mimo}
As explained in the previous section, thanks to the favorable properties of PMCW radar, the sidelobes caused by uncorrelated PN in the range profile are effectively attenuated \cite{b3}. Therefore, we propose to use range-domain MIMO to separate the mono- and bi-static signals. In that way, the mono- and bi-static signals will not interfere with each other.  

Two considerations should be noted: firstly, since the maximum unambiguous range is confined to \(\frac{L_c}{2}\), as detailed in Section~\ref{section A}, the range profile from range bin \(\frac{L_c}{2} + 1\) to \(L_c\) is discarded. Secondly, in the 2-radar network, for an equal separation of the range profile for the mono-static signal and the signals from another radar, at the transmitter side, the transmitted code sequence of Radar 2 is circularly shifted by \(\frac{L_c}{4}\). At the receiver side of both radars, the received code sequence does periodic correlation with the non-shifted code sequence.
Then, \eqref{eq:PN_ridge} and \eqref{eq:PN_ridge_los} should be modified as:
\begin{align} \label{eq:PN_ridge_circ}
     \nonumber & R_{bi,m}(p,n)= \sum_{l=0}^{L_c-1} (b(l)b(l-\tau_{d,bi}-\frac{L_c(m-1)}{4} + p))\\ \nonumber & e^{-j2\pi f_{D,bi}( (l+p)T_c + nT) }\\
     &  e^{j(\phi_{I1}((l+p)T_c + nT -\tau_{bi})-\phi_{I2}((l+p)T_c + nT))}\\ \nonumber
    & R_{los,m}(p,n) =  \sum_{l=0}^{L_c-1} (b(l)b(l- \tau_{d,los}-\frac{L_c(m-1)}{4} + p))\\  &
e^{j(\phi_{I1}((l+p)T_c + nT -\tau_{los})-\phi_{I2}((l+p)T_c + nT))} \label{eq:PN_ridge_los_circ}
\end{align}
where \(m = \{1;2\}\) is the index of the radar, and \(R_{bi,m}\) and \(R_{los,m}\) represent the signal from radar \(m\). This means we separate the range profile into two sections: the signals from Radar 1 are assigned to the range bin from 1 to \(\frac{L_c}{4}\), and the signals from Radar 2 are assigned to the range bin from \(\frac{L_c}{4} + 1\) to \(\frac{L_c}{2}\). This also indicates that the maximum unambiguous range will be reduced to \(\frac{L_c}{4}\) due to the range-domain MIMO.

If we want to extend to an M-radars network, the radar \(m\), \(1 \leq  m \leq M\), should be circularly shifted by \(\frac{L_c(m-1)}{2M}\). In this case, the range profile will be divided into M sections, one per radar, and the maximum unambiguous range is constrained to \(\frac{L_c}{2M}\).
\subsection{Uncorrelated PN in Doppler profile}
Since both the LOS signal and the bi-static signal originate from the same PLL, their PN is correlated. Given that the LOS signal Doppler frequency is zero, the phase variation of the LOS signal along slow time is solely determined by the PN samples. Therefore, we can extract the slow-time PN vector, $\xi(n)$, in the LOS range bin $\tau_{d,los} + \frac{L_c(m-1)}{4}$.
\begin{equation}
    \xi(n)=\angle R_{los,m}\left(\tau_{d,los}+\frac{L_c(m-1)}{4},n\right)
\end{equation}
where $\angle$ is the angle operator, then we can use the PN vector to compensate for the bi-static slow-time PN, as expressed by the following equation:
\begin{align} \label{eq:Solution}
    & R_{bi,m}\left(\tau_{d,bi}+\frac{L_c(m-1)}{4},n\right) e^{-j\xi(n)}\\
    \nonumber & = \sum_{l=0}^{L_c-1}e^{-j2\pi f_{D,bi}((l+\frac{L_c(m-1)}{4})T_c + nT+\tau_{bi} ) } \\ \nonumber & e^{j(\phi_{I2}((l+\frac{L_c(m-1)}{4})T_c + nT+\tau_{los})-\phi_{I2}((l+\frac{L_c(m-1)}{4})T_c + nT+\tau_{bi}))}\\
    \nonumber & = \sum_{l=0}^{L_c-1}e^{-j2\pi f_{D,bi}( (l+\frac{L_c(m-1)}{4})T_c + nT+\tau_{bi}) }e^{D_c(n)}
\end{align}
where the term $D_c(n)$ can be further derived as:
\begin{align}
    & D_c(n) = \sum_{k} j\alpha_k (\cos(x) - \cos(y)) \\
    \nonumber & = \sum_{k} j\alpha_k 2 \sin\left(\frac{y-x}{2}\right) \sin\left(\frac{x+y}{2}\right) \\
    \nonumber & = \sum_{k} \alpha_k  \sin\left(\pi f_k (\tau_{bi}-\tau_{los})\right) \\
    \nonumber & \left(e^{j(2\pi f_{k} ((l+\frac{L_c(m-1)}{4})T_c + nT + \frac{\tau_{bi}+\tau_{los}}{2}) + \theta_k)} \right. \\
    \nonumber & \left.
    -e^{-j(2\pi f_{k}((l+\frac{L_c(m-1)}{4})T_c + nT + \frac{\tau_{bi}+\tau_{los}}{2}) + \theta_k)}\right)
\end{align}
where \(x = 2\pi f_k ((l+\frac{L_c(m-1)}{4})T_c + nT+\tau_{los}) + \theta_k\) and \(y =  2\pi f_k ((l+\frac{L_c(m-1)}{4})T_c + nT+\tau_{bi}) + \theta_k\).  After compensation, similar to \eqref{eq:range_corr}, the bi-static PN can be attenuated by a factor of \(\sin\left(\pi f_k (\tau_{bi}-\tau_{los})\right)\). This attenuation is effective only when the condition \(f_{k}|\tau_{bi}-\tau_{los}| \leq \frac{1}{6}\) is satisfied. The main distinction is that here the effectiveness of PN attenuation depends on difference between LOS and bi-static delay while in \eqref{eq:range_corr} it only depends on the target location. 

Additionally, it is crucial to emphasize that if we follow the Doppler shift mitigation method from \cite{b1}, Doppler processing must precede range processing. However, in our approach to compensate for uncorrelated PN along the slow-time dimension, we first conduct range processing. To seamlessly integrate both solutions, we implement the Doppler shift mitigation first, followed by range processing. Subsequently, we perform an IDFT along the Doppler profile to extract the PN vector in the LOS range bin to compensate for the impact of uncorrelated phase noise along slow-time. Finally, we perform Doppler processing again to get a Range-Doppler map with low sidelobe and low noise floor. This Doppler shift mitigation technique can also be avoided if we use a code sequence which can maintain low range sidelobes in presence of Doppler frequency shift (e.g. P3 \cite{P3}).


\section{Numerical result}
In this section, we simulate a scenario where two PMCW radars are positioned side by side on a single platform, separated by a distance of 1m with no obstacles in between. Both radars utilize the same LO clock and are time-synchronized, enabling them to transmit PMCW waveforms simultaneously into free space.

\begin{figure}[] 
\centerline{\includegraphics[width =0.4\textwidth]{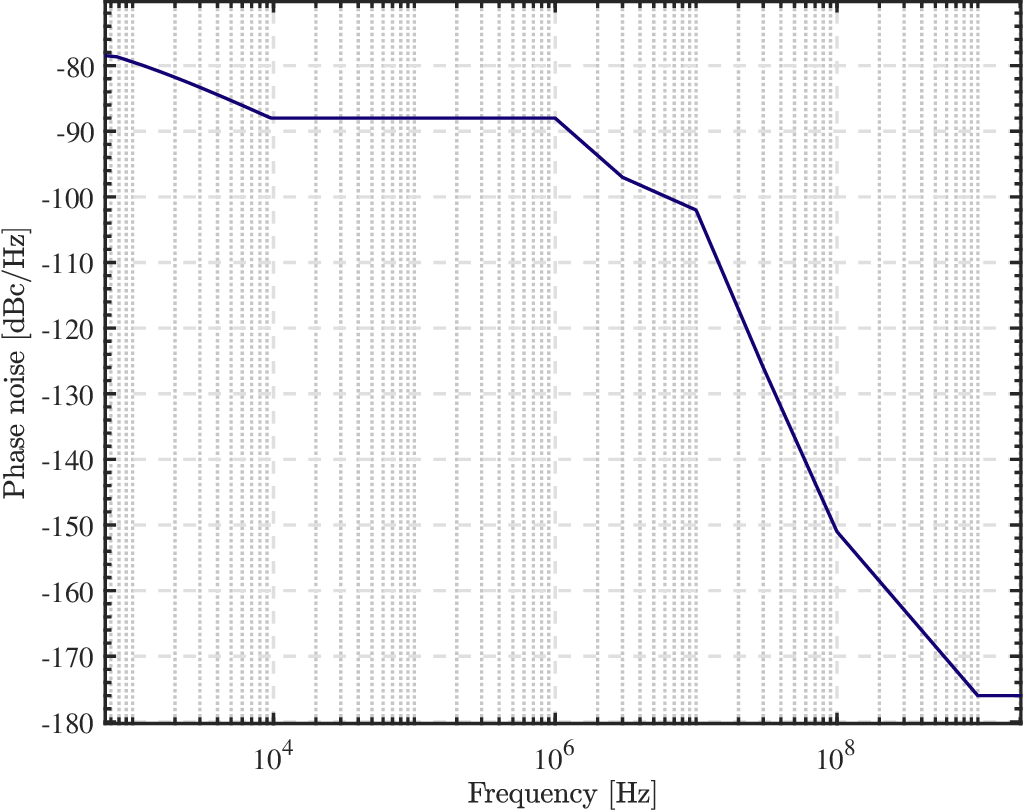}}
\caption{PN Power Spectral Density (PSD) used in this work to generate the PN samples}
\label{fig:psd_PN}
\end{figure}

Both radars share identical system configuration parameters, as illustrated in Table~\ref{table1}, experience uncorrelated PN with the same PSD (shown in Fig.~\ref{fig:psd_PN}), and exhibit identical antenna patterns. Additionally, since this paper mainly focuses on PN analysis, the impact of thermal noise is ignored.
\begin{table}[ht] 
\centering
\caption{System Configuration}
\begin{tabular}{|c|c|c|}
\hline
\textbf{Parameter} & \textbf{Value} & \textbf{Unit} \\
\hline
Transmit Power & 10 & dBm \\
\hline
Carrier Frequency & 79 & GHz \\
\hline
Chip Rate & 1 & ns \\
\hline
Sequence Length (Lc) & 504 &  \\
\hline
FFT Size (N) & 256 & \\
\hline
\end{tabular}
\label{table1}
\end{table}

Since the radars are placed side by side without obstacles, each radar can receive the LOS signal from the other radar at 90 degrees from boresight. The magnitude difference between the LOS path and the mono-static path can be defined by the ratio of the LOS path received power (defined by the Friis equation for one-way propagation) and the radar equation (for two-way propagation):
\begin{equation} \label{eq:bi-los_diff}
\frac{P_{\text{los}}}{P_{\text{mono}}} = \frac{4\pi G_{90}^2 R_{\text{mono}}^4}{G_t^2 R_{\text{los}}^2 \sigma^2}
\end{equation}
where  \(G_{90}\) represents the antenna gain at 90 degrees from boresight, and \(G_t\) is the antenna gain in the direction of the target. \(R_{\text{mono}}\) denotes the mono-static distance, \(R_{\text{los}}\) is the distance between the two radars, \(\sigma^2\) is the target RCS, \(P_{\text{los}}\) is the received power from the LOS path, and \(P_{\text{mono}}\) is the received power from the mono-static path.

\begin{figure}[] 
\centerline{\includegraphics[width=0.4\textwidth]{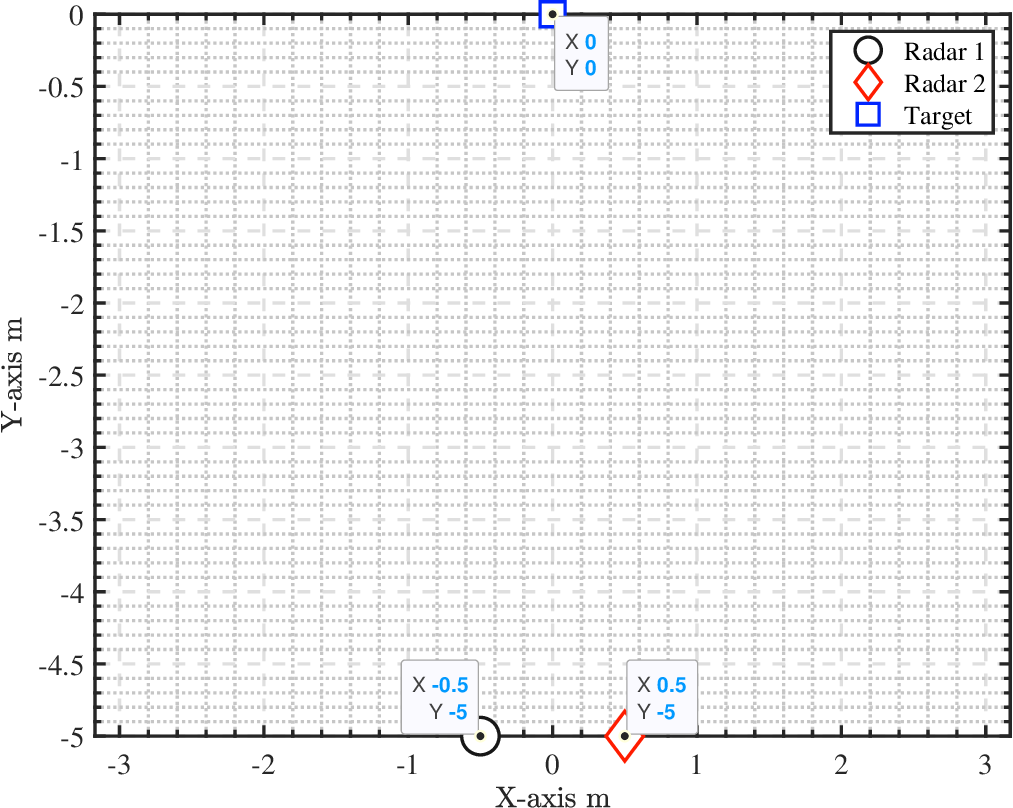}}
\caption{Radar and target position}
\label{fig:rd_position}
\end{figure}

Considering a static target located 5m from both radars as shown in Fig.~\ref{fig:rd_position}, with an RCS of 10 dBsm, the parameters of \eqref{eq:bi-los_diff} are given in Table.~\ref{los_table},
\begin{table}[ht] 
\centering
\caption{}
\begin{tabular}{|c|c|c|}
\hline
\textbf{Parameter} & \textbf{Value} & \textbf{Unit} \\
\hline
$G_{90}$ & $-7$ & dB \\
\hline
$G_t$ & $10$ & dB \\
\hline
$R_{\text{mono}}$ & $5$ & m \\
\hline
$R_{\text{los}}$ & $1$ & m \\
\hline
$\sigma^2$ & $10$ & dBsm \\
\hline
\end{tabular}
\label{los_table}
\end{table}
where the value of the antenna gain is determined based on the specifications of the antenna pattern of commercial radars (\cite{TI}, Fig. 11).
Applying (29), the LOS signal is found to be 5 dB weaker than the mono-static signal.
\begin{figure}[] 
\centerline{\includegraphics[width=0.5\textwidth]{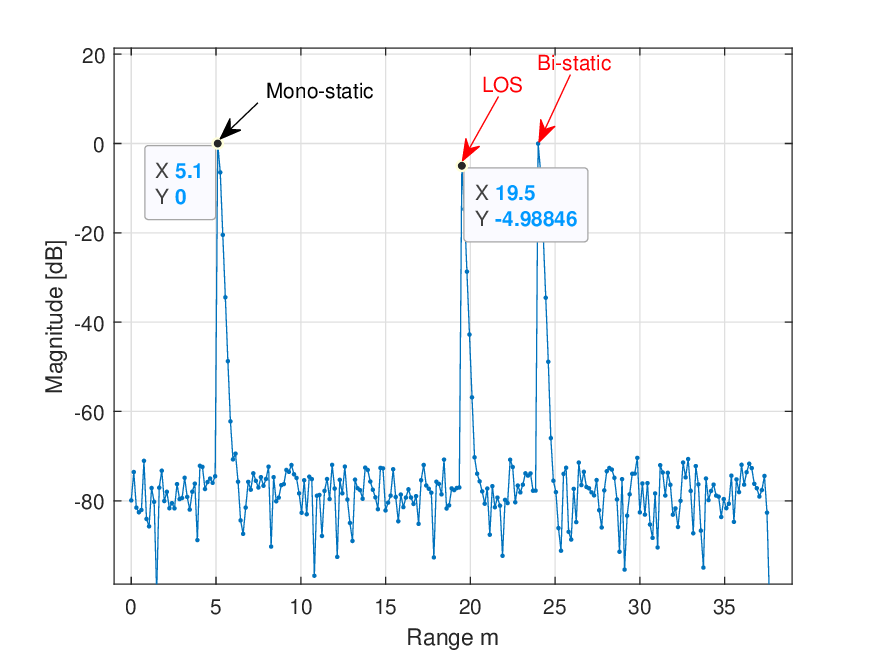}}
\caption{Radar 1 range profile (DC Doppler bin, normalized to 0 dB)}
\label{fig:radar1_range_profile}
\end{figure}
As shown in Fig.~\ref{fig:radar1_range_profile}, the sidelobes caused by uncorrelated PN are observed to be attenuated in the range profile. The mono-static signal is assigned to the range profile from range bin 1 to \(\frac{L_c}{4}\), while the bi-static signal and LOS signal from radar 2 are assigned to the range profile from \(\frac{L_c}{4} + 1\) to \(\frac{L_c}{2}\), as explained in Section \ref{range_mimo}.
\begin{figure}[]
    \centering
    \begin{subfigure}[b]{0.4\textwidth}
        \includegraphics[width=\textwidth]{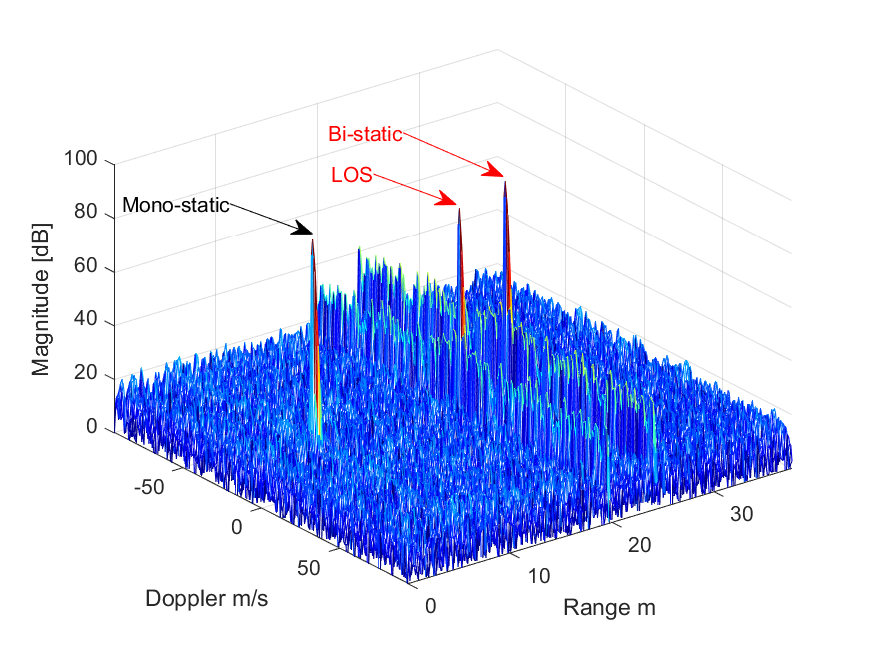}
        \caption{Original}\label{fig:RDM}
    \end{subfigure}
    \hspace{1cm}
    \begin{subfigure}[b]{0.4\textwidth}
        \includegraphics[width=\textwidth]{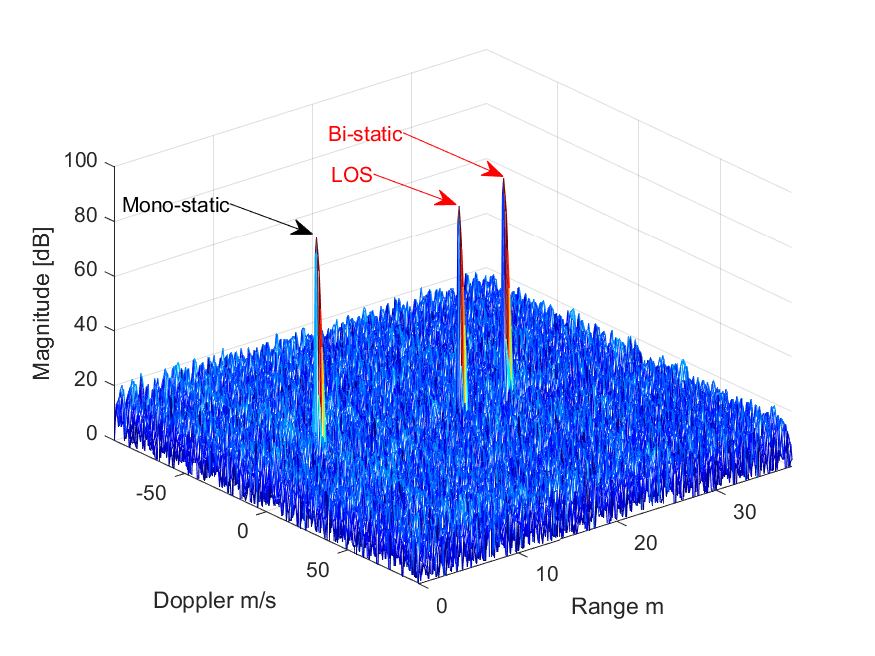}
        \caption{After compensation}\label{fig:RDM_COMP}
    \end{subfigure}
    \caption{Range-Doppler Maps of Radar 1}
\end{figure}
Fig.~\ref{fig:RDM} shows that the Bi-static signal and LOS signal produce a ridge along the Doppler profile due to the uncorrelated PN, as expected and explained in \eqref{eq:PN_ridge_bi_rangebin} and \eqref{eq:PN_ridge_los_rangebin}. To attenuate the ridge, after the periodic correlation processing, we can locate the LOS in the range profile, extract the PN vector, and apply the method described in \eqref{eq:Solution} to both the LOS range bin and the bi-static range bin.
After the compensation, the ridge of the bi-static signal and LOS signal has been removed, as shown in Fig.~\ref{fig:RDM_COMP}.


\section{Conclusion}
This paper proposes a solution for building a multi-static radar network by leveraging the PN robustness of PMCW to attenuate interferences between multi- and mono-static paths. We present a comprehensive mathematical model that describes the impact of both correlated and uncorrelated PN along the slow-time dimension. Additionally, we introduce a solution to mitigate the influence of uncorrelated PN in the Doppler dimension.

As a result, the proposed radar network solution provides a low noise floor even in the presence of uncorrelated PN, eliminating the need for sharing carrier frequency signals among radars or a calibration target. 

This research offers a practical approach to enhance radar performance in multi-static radar networks.

\section*{Acknowledgment}
This work has been supported by the SNS JU project SUNRISE-6G (Grant Agreement No. 101139257) under Horizon Europe research and innovation programme.


\vspace{12pt}


\begin{thebibliography}{00}
\bibitem{b8} K. Siddiq, M. K. Hobden, S. R. Pennock and R. J. Watson, "Phase Noise in FMCW Radar Systems," in IEEE Transactions on Aerospace and Electronic Systems, vol. 55, no. 1, pp. 70-81, Feb. 2019.
\bibitem{b9} M. Gottinger et al., "Coherent Automotive Radar Networks: The Next Generation of Radar-Based Imaging and Mapping," in IEEE Journal of Microwaves, vol. 1, no. 1, pp. 149-163, Jan. 2021.
\bibitem{b10} B. Meinecke, M. Steiner, J. Schlichenmaier and C. Waldschmidt, "Instantaneous Target Velocity Estimation Using a Network of a Radar and Repeater Elements," 2019 16th European Radar Conference (EuRAD), Paris, France, 2019, pp. 241-244.

\bibitem{b11} D. Werbunat, B. Schweizer, B. Meinecke, R. Michev, J. Hasch and C. Waldschmidt, "Ghost-Target Suppression in Coherent Radar Networks," 2021 18th European Radar Conference (EuRAD), London, United Kingdom, 2022, pp. 54-57.
\bibitem{b1} S. Xu and A. Yarovoy, "Doppler Shifts Mitigation for PMCW Signals," 2019 International Radar Conference (RADAR), Toulon, France, 2019, pp. 1-5.

\bibitem{phase_noise} A. Dürr, D. Böhm, D. Schwarz, S. Häfner, R. Thomä and C. Waldschmidt, "Coherent Measurements of a Multistatic MIMO Radar Network With Phase Noise Optimized Non-Coherent Signal Synthesis," in IEEE Journal of Microwaves, vol. 2, no. 2, pp. 239-252, April 2022.
\bibitem{range_corr} M. C. Budge and M. P. Burt, "Range correlation effects on phase and amplitude noise," Proceedings of Southeastcon '93, Charlotte, NC, USA, 1993, pp. 5 p.-.
\bibitem{b3} M. Bauduin and A. Bourdoux, "Impact of Phase Noise on FMCW and PMCW Radars," 2023 IEEE Radar Conference (RadarConf23), San Antonio, TX, USA, 2023, pp. 1-6.
\bibitem{b12} H. C. Yildirim, M. Bauduin, A. Bourdoux and F. Horlin, "Impact of Phase Noise on Mutual Interference of FMCW and PMCW Automotive Radars," 2019 16th European Radar Conference (EuRAD), Paris, France, 2019, pp. 181-184.
\bibitem{b13} A. Dürr, R. Kramer, D. Schwarz, M. Geiger and C. Waldschmidt, "Calibration-Based Phase Coherence of Incoherent and Quasi-Coherent 160-GHz MIMO Radars," in IEEE Transactions on Microwave Theory and Techniques, vol. 68, no. 7, pp. 2768-2778, July 2020.
\bibitem{b2} F. Horlin and A. Bourdoux. Digital Compensation for Analog Front-Ends: A New Approach to Wireless Transceiver Design. Wiley and Sons, 2008.


\bibitem{b6} W. Van Thillo, P. Gioffré, V. Giannini, D. Guermandi, S. Brebels and A. Bourdoux, "Almost perfect auto-correlation sequences for binary phase-modulated continuous wave radar," 2013 European Radar Conference, Nuremberg, Germany, 2013, pp. 491-494.

\bibitem{TI} Texas Instruments. AWR1443, AWR1243 Evaluation Module (AWR1443BOOST, AWR1243BOOST) mmWave Sensing Solution.\\ 
\url{https://www.ti.com/lit/ug/swru507c/swru507c.pdf?ts=1715724817901}.

\bibitem{P3} M. Bauduin and A. Bourdoux, "Reducing Range Sidelobes and Ghost Targets in PMCW Radars With $\pi$/K-Zadoff Code Sequences," in IEEE Transactions on Radar Systems, vol. 1, pp. 646-656, 2023.

\end{thebibliography}
\end{document}